\RequirePackage{fix-cm}
\documentclass[smallextended]{svjour3}       % onecolumn (second format)
\smartqed  % flush right qed marks, e.g. at end of proof
\usepackage{graphicx}
\usepackage[misc]{ifsym} % for envelope icon
\RequirePackage[colorlinks=true, allcolors=blue]{hyperref}
\usepackage{booktabs}

\graphicspath{{graphics/}}

\begin{document}

% \title{Discussion Cascades of Conspiracy and Science Information Online}

% \title{Cascades of Conspiracy and Science Discussion Threads}

% \title{A Large-scale Analysis of Online Conspiracy and Science Discussion Cascades}

\title{Conspiracy vs Science: A Large-scale Analysis of Online Discussion Cascades}

% \titlerunning{A Large-scale Analysis of Online Discussion Cascades}        % if too long for running head

\authorrunning{Yafei Zhang et al.}

\author{Yafei Zhang$^{1,2}$         \and
        Lin Wang$^{1}$            \and
        Jonathan J. H. Zhu$^{2}$  \and
        Xiaofan Wang$^{1,3}$ %etc.
}

\institute{
            Yafei Zhang \at
              \email{yflyzhang@gmail.com}           %  \\
           \and
           Lin Wang \at
              \email{wanglin@sjtu.edu.cn}
           \and
           Jonathan J. H. Zhu (\Letter) \at
              \email{j.zhu@cityu.edu.hk}
           \and
           Xiaofan Wang (\Letter) \at
              \email{xfwang@sjtu.edu.cn}
            \and
        $1.$ Department of Automation, Shanghai Jiao Tong University, and Key Laboratory of System Control and Information Processing, Ministry of Education of China, Shanghai 200240, China\\
        $2.$ Department of Media and Communication, and School of Data Science, City University of Hong Kong, Hong Kong S.A.R., China\\
        $3.$ Department of Automation, Shanghai University, Shanghai 200444, China\\
}


% ================================
% \title{Insert your title here%\thanks{Grants or other notes
% %about the article that should go on the front page should be
% %placed here. General acknowledgments should be placed at the end of the article.}
% }
% \subtitle{Do you have a subtitle?\\ If so, write it here}


% \author{First Author         \and
%         Second Author %etc.
% }


% \institute{F. Author \at
%               first address \\
%               Tel.: +123-45-678910\\
%               Fax: +123-45-678910\\
%               \email{fauthor@example.com}           %  \\
% %             \emph{Present address:} of F. Author  %  if needed
%               second address
% }


\maketitle

\begin{abstract}

With the emergence and rapid proliferation of social media platforms and social networking sites, recent years have witnessed a surge of 
misinformation spreading in our daily life.
Drawing on a large-scale dataset which covers more than 1.4M posts and 18M comments
from an online social media platform, we investigate
the propagation of 
two distinct narratives--(i) conspiracy information, whose claims are generally unsubstantiated and thus 
referred as misinformation to some extent,
and (ii) scientific information, whose origins are generally readily identifiable and verifiable.
We find that
conspiracy cascades tend to propagate in a multigenerational branching process
whereas
science cascades are more likely to grow in a breadth-first manner.
Specifically,
conspiracy information triggers larger cascades, 
% involves more users, spreads deeper, 
involves more users and generations, 
persists longer, and is more viral and bursty than science information.
% 
% We also find that
Content analysis reveals that
conspiracy cascades contain more 
negative words and emotional words which convey
anger, fear, disgust, surprise and trust.
% 
% The analysis also shows that
We also find that
% conspiracy and science cascades focus on quite different topic landscapes, where
conspiracy cascades are much more concerned with political and controversial topics.
After applying machine learning models, we achieve an AUC score of nearly 90\% in 
% identifying whether a given post belongs to conspiracy or science narratives using the features obtained above.
discriminating conspiracy from science narratives using the constructed features.

We further investigate 
user's role 
during the growth of cascades.
In contrast with previous assumption that misinformation is primarily driven by a small set of users,
% distribute and propagate by
we find that conspiracy cascades 
are more likely to be controlled by a broader set of users than science cascades,
% raising 
imposing
new challenges 
on
% to 
the management of misinformation.
% *******************
Although political affinity 
% of users 
is thought to affect the consumption of misinformation,
there is very 
% few 
little
evidence that political orientation of the information source plays a role during the propagation of conspiracy information;
Instead, we find that 
% low-credibility 
conspiracy
information from media outlets with left or right orientation triggers smaller cascades and is less viral than information 
from online social media platforms (e.g., Twitter and Imgur)
whose political orientations are unclear.
Our study 
provides complementing evidence to current misinformation research
and has practical policy implications to
stem the propagation and mitigate the influence of misinformation online.

\keywords{Misinformation \and
Conspiracy \and
Information cascades \and
Online community \and Political affinity}

% Insert your abstract here. Include keywords, PACS and mathematical
% subject classification numbers as needed.
% \keywords{First keyword \and Second keyword \and More}
\end{abstract}

\section{Introduction}
\label{intro}
% Your text comes here. Separate text sections with

The emergence of digital technologies 
such as e-mail, online social networks and instant messages,
% social media platforms and social networking sites 
has dramatically shifted the way we get and consume information
and provides an unprecedented opportunity to novel investigations of the information aggregation in networks on a large scale
\cite{anderson2015global,cheng2014can,goel2012structure,liang2018broadcast,pei2015exploring,qiu2016lifecycle}.
Although
% these online social systems are
new social technologies are
thought to foster the aggregation and consumption of news,
% but it also foster the diffusion of rumors and .
they may also
contribute to fueling the dissemination of 
rumors or misinformation
in today's society.

Misinformation research
has attracted increasing attention in recent years,
% mainly due to the 2016 
primarily under the political context such as the US presidential election \cite{allcott2017social,bovet2019influence,grinberg2019fake,guess2018selective,lazer2018science}
and partisanship \cite{guess2019less,pennycook2019fighting},
% 
% but it also has been documented in a variety of topics, such as the 
but it also has arisen in a variety of topics such as
extreme events \cite{huang2015connected,jones2017distress,starbird2014rumors},
social bots \cite{shao2018spread,stella2018bots}
and
rumor detection \cite{almaatouq2016if,friggeri2014rumor,qazvinian2011rumor,shu2017fake}.
More recently, 
the growing surge of misinformation about COVID-19 disease
even poses risks to global health\footnote{\url{https://time.com/5811939/un-chief-coronavirus-misinformation}.}.
Misinformation is able to
exacerbate distress during terrorist attack,
alter public opinion,
affect election outcome
and
create a climate of mistrust,
as such it's of prime significance to understand the propagation and social origins of it \cite{del2016spreading,vosoughi2018spread}.

As more and more people are reading news online
\cite{gottfried2017americans,newman2017reuters},
online social media sites
yield invaluable sources for
probing the underlying mechanisms behind the formation and propagation of misinformation
due to the ready availability of large-scale datasets \cite{mocanu2015collective}.
Several 
valuable efforts have been made 
to the investigation of
fake news or misinformation
\cite{bovet2019influence,grinberg2019fake,vosoughi2018spread},
but most of them have focused largely, if not exclusively, on 
user-based platforms such as Facebook or Twitter in the main analysis.
While for
interest- or community-based platforms such as StackExchange or Reddit,
there is still an immense shortage of investigation on the circulation of misinformation on such platforms,
especially for communities or interest groups 
where misinformed or unsubstantiated messages are chronically circulated.
Moreover,
user engagement,
which measures the intensity of user involvement during the propagation of misinformation,
has been largely overlooked in previous studies \cite{allcott2017social,bovet2019influence,grinberg2019fake,guess2019less,guess2018selective}.
In addition,
although people with different political orientations are inclined to consume specific kinds of misinformation \cite{guess2018selective},
whether political orientation of the information source matters in the circulation of misinformation in particular communities
is still poorly understood.

% In this work, 
Leveraging data from two Reddit\footnote{\url{https://www.reddit.com}.} communities or subreddits--r/conspiracy and r/science,
we conduct a large-scale quantitative analysis
% which aims to give insights
through extensive comparison between 
conspiracy and science discussion threads
in respective community.
In this work, we follow the practice in \cite{del2016spreading},
where 
conspiracy theories are considered unsubstantiated due to their lack of verifiability
while science narratives are generally identifiable and verifiable.
Note that, we don't claim that all conspiracy narratives in r/conspiracy are fake;
Instead, we refer them as misinformation or low-credibility contents
due to their nature of uncertainty and incitement \cite{del2016spreading}.
For example, 
a lot of users in r/conspiracy doubt about the veracity of
the Apollo moon landing project
which is a widely used case by conspiracists.
If these 
misinformed or unsubstantiated messages
are proliferating
over the Internet,
it would be dangerous for the 
form of public opinion in the real world.
On the contrary, r/science is a community to share and discuss scientific news and advances whose origins are generally identifiable and verifiable.
For instance,
many conspiracy theories claim that 
the COVID-19 coronavirus, which is causing the pandemic, 
is created in a lab,
while scientific analysis would suggest that 
the coronavirus is not a purposefully manipulated virus.
Also note that we focus on the possibility of verification of the information rather than the quality of the information.
Therefore, 
the systematic comparison 
between conspiracy and science discussion cascades 
should be able to
yield
a clear understanding of the patterns
that misinformation differs from substantiated information.

\vspace{4mm}
\noindent \textbf{The Present Work.}
In this work,
we first analyze the cascade dynamics of conspiracy and science threads in respective community.
Generally,
a discussion thread consists of a post as well as comments under the post arranged in a hierarchical way, 
and
can be naturally mapped into a cascade tree
with
the post acting
% acts
as the root and
the hierarchical comments
forming
% form 
the branches and leaves in a sequential order.
We mainly consider the differences between conspiracy and science cascades
in terms of 
several crucial structural and temporal properties, including cascade size, depth, virality and lifetime.
Our analysis suggests that
% We find that 
% generally speaking,
% 
conspiracy information tends to trigger larger cascades, get more individuals involved, 
propagate further,
% more generations
survive longer, and is more viral and bursty than science information.
Our findings are consistent with previous studies \cite{del2016spreading,vosoughi2018spread}
in terms of these structural and temporal properties.
%  "our findings agree with…"

We also find that 
conspiracy cascades tend to circulate in a multigenerational branching manner,
while science cascades are likely to grow in a breadth-first way
where science cascades attract more users at each generation or depth 
compared with conspiracy cascades.
% 
% Furthermore,
In general,
it takes conspiracy cascades less time to reach the same depth as science cascades, 
but it will cost much more time for them to grow into 
larger cascades (e.g., $\geq$ 20).
It is also important to note that
there are more science cases than conspiracy cases 
among the largest cascades (e.g., $\geq$ 1,000),
which is 
not found
in user-based platforms \cite{del2016spreading,vosoughi2018spread}.

After applying content analysis,
we find that 
compared with science cascades,
conspiracy cascades 
are much richer in sentiment
and 
convey more emotions of anger, fear, disgust, surprise and trust in their contents.
Some of the findings are also consistent with previous study \cite{vosoughi2018spread}.
Topic modeling further reveals that
conspiracy narratives tend to
pay more attention to political and controversial topics.
In light of 
the disparity between conspiracy and science cascades in terms of 
dynamic, emotion and topic features,
we implement a serial of classification tasks to distinguish conspiracy narratives from science ones.
We achieve good classification performance--an AUC score of nearly 90\%--using all three sets of features in ten random trials where 80\% of the data are assigned as the training set while the rest 20\% of the data as the test set.

To study 
the intensity of user engagement 
during the propagation of conspiracy or science narratives,
two comment networks are constructed
based on the explicit comment paths between users in each community.
After applying
percolation process on comment networks,
we find that 
conspiracy comment network is less concentrated on focal nodes compared with science comment network,
suggesting that
conspiracy cascades are less likely to be
driven by a few set of users than expected.

We further explore 
the relationship between political orientation and circulation of misinformation.
Leveraging
a fact-checking list of media outlets and the URLs embedded in the contents,
we are able to assign conspiracy narratives into different political subgroups,
such as left, left leaning, center, right leaning and right.
However, there is very 
% few 
little
evidence that 
political orientation contributes to
the circulation of conspiracy information
as the size, depth and virality of cascades from different political subgroups
are either smaller than or nearly equal to the general case.
Instead,
we find that 
conspiracy information
from online social media platforms like Twitter and Imgur generally induces larger cascades, propagates further and is more viral than the general.

Our work presents important findings complementing previous misinformation research
and could offer valid insights for 
the fight with misinformation under current situation.

\section{Related work}

There are mainly four 
lines of research related to our current work:

\subsection{Misinformation}
Recent years have brought the term ``misinformation'' into broad circulation due to the prevalence of fake news in online social media platforms.
The use of the term misinformation could vary drastically from the view of researchers from different backgrounds \cite{ruths2019misinformation}.
Here we follow the practice in \cite{bessi2015science,del2016spreading}
where conspiracy narratives are treated as misinformation
% ,and refer conspiracy narratives as low-credibility contents 
due to their lack of verifiability.

The circulation of 
misinformation over the Internet has great potential to
endanger the democratic election, 
increase public distress during terror attacks,
induce misallocation of resources after natural disasters 
and damage the credibility of regulatory agencies
\cite{Aral858,jones2017distress,lazer2018science,vosoughi2018spread}.
Although there has been a surge of misinformation research recently, most of them
% lack of 
either use small data samples \cite{allcott2017social,guess2018selective} or
only focus on
user-based platforms \cite{bovet2019influence,vosoughi2018spread}.
As a complement,
our study is based on two interest-based communities in Reddit, 
which are homes for sharing and discussing conspiracy and scientific narratives.
The dataset is thought to
cover comprehensive historical data of discussion threads in each community with a time span of more than ten years.

\subsection{Information cascades}
The growing availability of large scale digital traces,
along with the development of computational techniques to analyze them,
has fostered
extensive empirical investigations on information cascades.
Recently,
there has been a surge of interests in terms of 
diverse information cascades,
such as 
reshare of photos \cite{cheng2014can},
retweet of posts \cite{del2016spreading,liang2018broadcast,vosoughi2018spread},
adoption of new products \cite{anderson2015global,goel2012structure}
and
formation of social groups \cite{qiu2016lifecycle}.
For example, in \cite{del2016spreading} the successive sharing of  news in Facebook is conceptualized as information cascade.

% Conversion or 
Discussion threads,
% unlike aforementioned actions,
which involve elaborated conversational interactions between users,
can also be characterized as cascades \cite{gomez2011modeling,gomez2013likelihood,kumar2010dynamics}.
Our study focuses on discussion threads, where the successive `comment-to' relations indicate the information flows.
From the view of network analysis,
discussion threads can be naturally represented by tree structure,
with nodes 
% represent 
representing
a post and the comments under it and edges 
% represent 
representing
comment-to actions \cite{gomez2011modeling,kumar2010dynamics,medvedev2019modelling}.
We would apply network analysis in our study to 
depict 
several crucial 
structural properties from the constructed discussion cascades.

\subsection{Text mining}
Text mining provides technical support for further investigation of contents
beyond merely structural or temporal properties of cascades.
A number of studies have adopted the
word-emotion association lexicon to make automatic evaluation of sentiments and emotions embedded in texts \cite{brady2017emotion,cheng2017anyone,romero2016social,vosoughi2018spread}.
In addition, topic modeling is another widely used method to quantify the underlying topic concerns 
that characterize a set of documents
\cite{blei2003latent,guan2016segmenting,park2018strength,singer2017why,vosoughi2018spread,zhang2017using}.
For example, 
there are significant emotional gaps between 
true and false news \cite{vosoughi2018spread}, 
and users' topic concerns are also shown to be helpful to
predict their adoption behaviors \cite{guan2016segmenting}.
In this study, we follow these practices to elicit sentiments, emotions and topics that are conveyed in the narratives,
and explore to what extent these features help
to discriminate conspiracy narratives from scientific ones.

\subsection{Political affinity}
Political affinity is thought to 
% be able to 
affect the consumption and spread of information.
As documented in the literature,
% it's thought that
Trump supporters are likely to visit the most fake news websites \cite{guess2018selective},
and people with the right political affinity 
are more likely to share fake news on Twitter
than people on the left do so \cite{grinberg2019fake}.
It also has been shown that
traditional center and left leaning news spreaders
have stronger influence on the activity of Clinton supporters than that of Trump supporters during the 2016
US presidential election \cite{bovet2019influence}.
These analyses have focused exclusively on the political affinity of news spreaders instead of the political orientation of the information source.
Here in our study, we aim to investigate the role of the political orientation (e.g., left, center or right) 
of conspiracy narratives
plays during 
the propagation of misinformed cascades.

% \noindent

\section{Data}

We use data from Reddit\footnote{The raw data we used in this study are acquired from and publicly available at \url{https://files.pushshift.io/reddit}.}, a social news and discussion website, which ranks in the top 20 websites in the world according to the traffic.
The dataset
covers

\begin{figure*}[!hb]
    \centering
    \includegraphics[scale=.5]{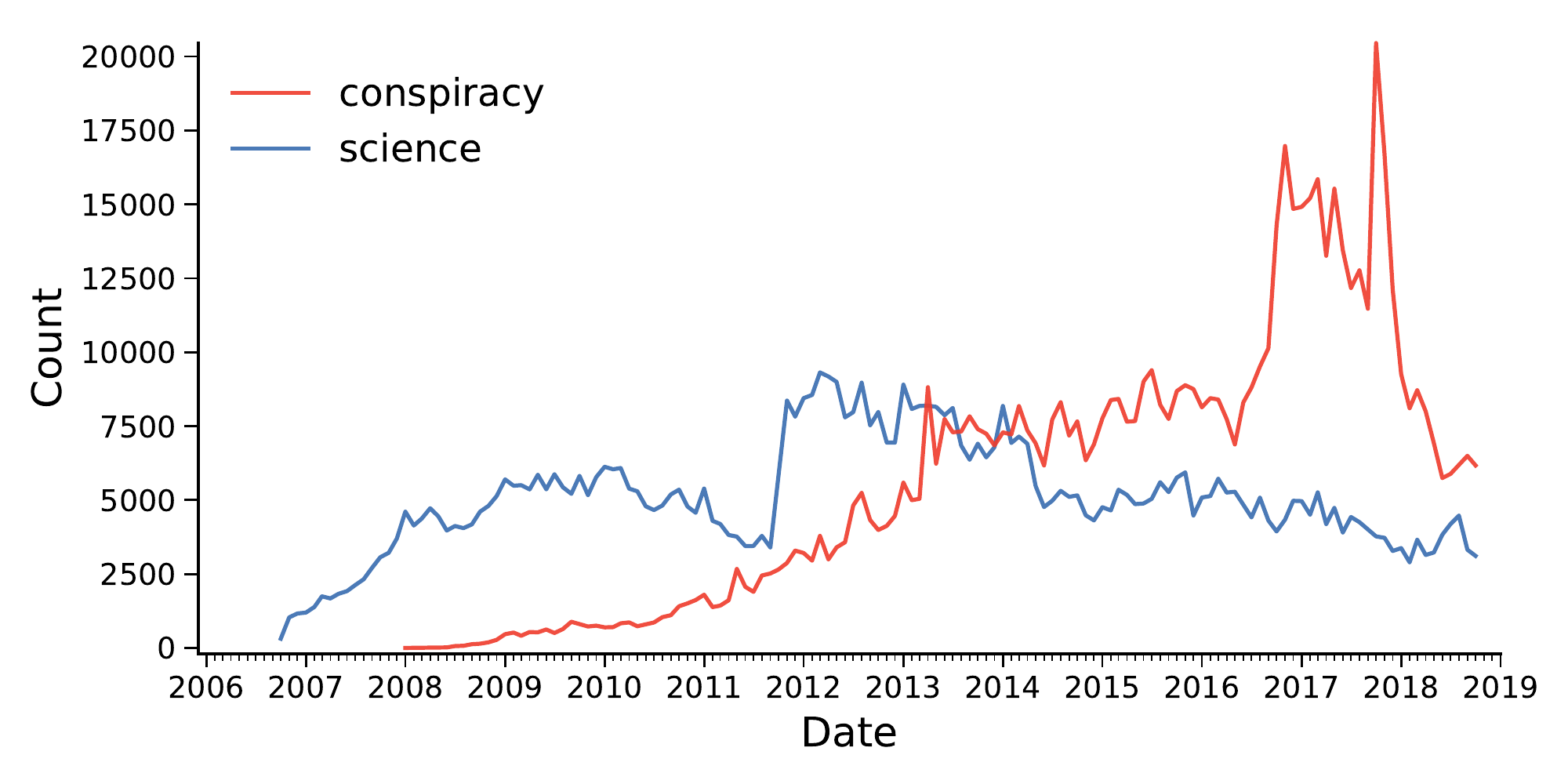}
    \caption{
    \textsf{
    % \small
    \footnotesize
    \textbf{Monthly cascade count (2006-2018).}
    }}
    \label{fig:monthly_cnt}
\end{figure*}

\noindent
more than 1.4M posts and 18M comments
under two distinct 
online communities or subreddits (r/science and r/conspiracy) 
over a time span of more than ten years (until October 2018).
Figure \ref{fig:monthly_cnt} shows the monthly cascade count for both conspiracy and science posts.
Clearly, we see a surge of conspiracy cascades during the US presidential election (e.g., the 2016 presidential election), which 
may indicate
that the prevalence of misinformation is likely to be catalyzed by political events.

\begin{figure*}[!t]
% \begin{figure}[!htbp]
    \centering
    \includegraphics[scale=.7]{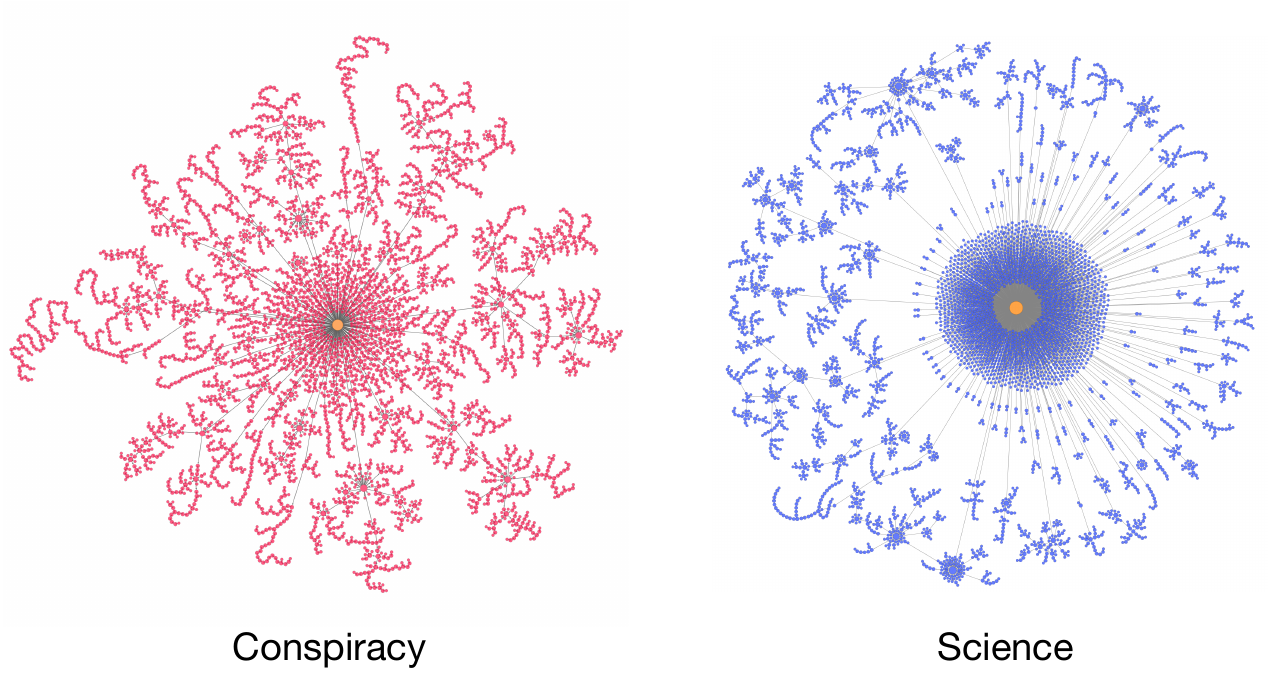}
    \caption{
    \textsf{
    % \small
    \footnotesize
    \textbf{Two cascade examples from conspiracy and science subreddits.}
    % , respectively.
    }
    }
    \label{fig:cas_exs}
\end{figure*}

Figure \ref{fig:cas_exs} illustrates two examples of conspiracy and science cascades, where the source (i.e., the post) of each cascade is highlighted in orange and the edge indicates the comment path (e.g., $A \rightarrow B$ means that $B$ is a successor or child of $A$).
Intuitively, conspiracy cascade tends to comprise a 
multi-step propagation
process where any one node directly 
connects
only a few others,
while science cascade corresponds to a broadcast-dominated 
% diffusion process
propagation process
where a small set of nodes directly connects a large proportion of nodes.
We will elaborate on this point in the next section.

% \newpage

\begin{figure}[!t]
% \begin{figure}[!htbp]
    \centering
    \includegraphics[scale=.65]{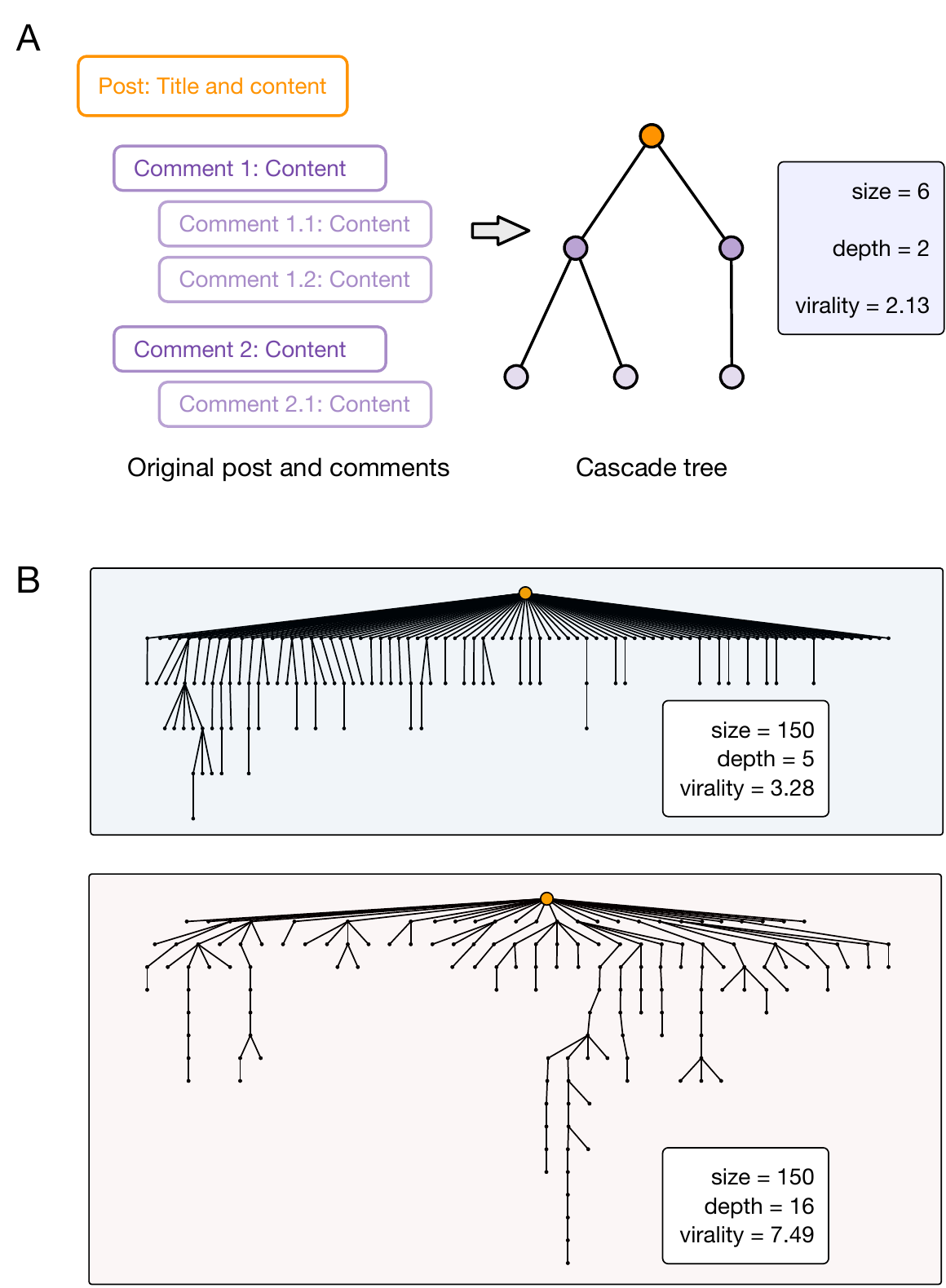}
    \caption{
    \textsf{
    % \small
    \footnotesize
    \textbf{Cascade tree.}
    (\textit{A}) Post and hierarchical comments to a cascade tree.
    % (\textit{B}) An example cascade of conspiracy news.
    % \textbf{B} Conspiracy.
    (\textit{B}) Two cascades with the same size but varying depth.
    }
    }
    \label{fig:comment2cascade}
\end{figure}

\section{Cascade dynamics}

\subsection{Primary indicators}

Figure \ref{fig:comment2cascade}A delineates the process of how a post (colored in orange) and the hierarchical comments (colored in purple) under it are transformed into a cascade tree, where the root of the tree is colored in orange.
The primary indicators of the cascade tree 
% we considered here 
that we are interested in
are described as follows:

\begin{itemize}
\item {\verb|size|}: number of nodes in a cascade tree (including the root).
\item {\verb|unique users|}: number of unique users involved in a cascade tree.
\item{\verb|depth|}: maximum number of hops from the root of a cascade tree. 
\item{\verb|virality|}: the average path length between all pairs of nodes in a cascade tree (also known as ``Wiener index'' or ``structural virality")\cite{goel2015structural,goel2012structure}.
% For a given diffusion tree $T$, the structural virality $V_T$ is defined as the average distance between all pairs of nodes in $T$.
\item{\verb|lifetime|}: time span (in minutes) of the last comment under a post since the post is published.
\item{\verb|burstiness|}: a measure which interpolates between comments that come in a periodic manner and those come in a bursty manner\cite{goh2008burstiness}.
Formally, for a Poisson process,
% Given a set of signals, 
where each discrete signal records the moment when the event occurs,
the interevent time, $\tau$, is defined as the time difference between two consecutive events.
% 
% In this case the interevent time, τ, between two consecutive events follows an exponential distribution
% 
The burstiness parameter is then defined as
% 
% $B \equiv \frac{\left(\sigma_{\tau} / m_{\tau}-1\right)}{\left(\sigma \tau / m_{\tau}+1\right)}=\frac{\left(\sigma_{\tau}-m_{\tau}\right)}{\left(\sigma_{\tau}+m_{\tau}\right)}$
% 
$B=\frac{\left(\sigma_{\tau}-m_{\tau}\right)}{\left(\sigma_{\tau}+m_{\tau}\right)}$,
where $m_{\tau}$ is the mean of $\tau$
and $\sigma_{\tau}$ is the standard deviation of 
$\tau$.
% and $\sigma_{\tau}$ are the mean and the standard deviation of 
% $\tau$.
\end{itemize}

\noindent These indicators form a set of crucial structural and temporal features that characterize a cascade.
Note that,
for cascade trees with only one node (i.e., the root), their depth, virality and lifetime are defined as 0.

According to the definition above,
% As shown in Figure \ref{fig:comment2cascade}A, 
the example cascade in Figure \ref{fig:comment2cascade}A has a size of 6, a depth of 2 and a virality measure of 2.13.
Figure \ref{fig:comment2cascade}B further illustrates two example cascades from the data, which have the same size but varying depth and virality.
The upper cascade comes from the science community while the lower cascade comes from the conspiracy community.
Both of the cascade trees have 150 nodes, but 
most of the nodes are immediately or intermediately connected with the root (i.e., one or two hop away from the root) for the upper one while a large proportion of nodes are 
% more than 
beyond
one hop away from the root for the lower one.
As expected, the virality of the lower cascade tree is much larger than that of the upper one (7.49 versus 3.28).
In other words, the lower cascade is more viral 
% or less broadcast 
than the upper cascade.

\begin{figure*}[!b]
    \centering
    \includegraphics[scale=.5]{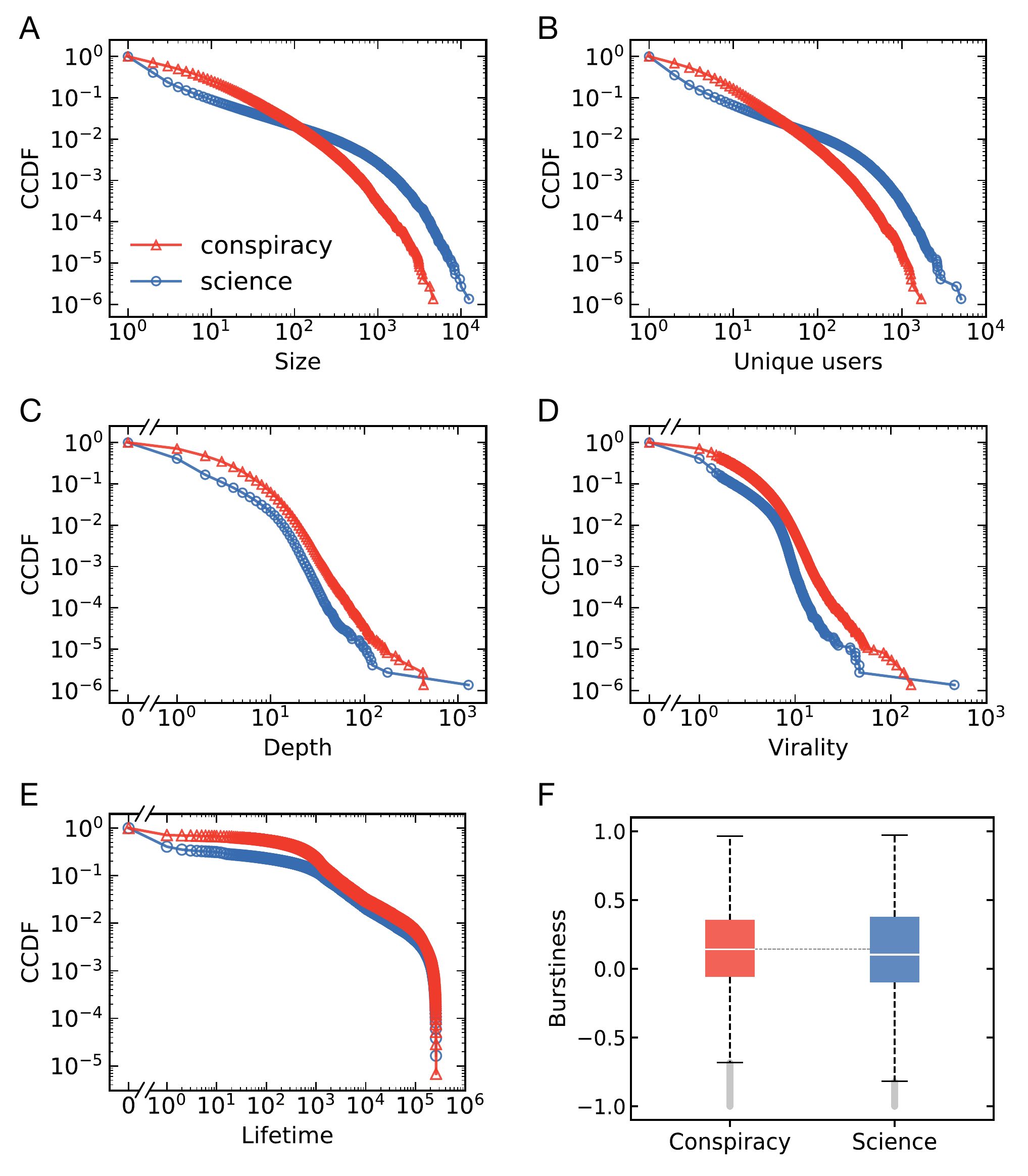}
    \caption{
    \textsf{
    % \small
    \footnotesize
    \textbf{Cascade dynamics of conspiracy and science information.
    % Distributions of cascade size, unique users, depth, virality, lifetime and burstiness.
    }
    (\textit{A}) CCDF (Complementary Cumulative Distribution Function) of cascade size.
    (\textit{B}) CCDF of the number of unique users.
    (\textit{C}) CCDF of cascade depth.
    (\textit{D}) CCDF of virality.
    (\textit{E}) CCDF of cascade lifetime (in minutes).
    (\textit{F}) Burstiness of cascades.
    }}
    \label{fig:spread_diff}
\end{figure*}

\subsection{Cascade dynamics}

% According to 
In light of
the introduced structural and temporal indicators of a cascade, 
we analyze the dynamics of cascades induced by conspiracy and science information.
To construct the cascade trees, we simply remove comments imposed by the Reddit bot--\textit{u/AutoModerator}.
% u/AutoModerator
% 
% We find that,
Analysis reveals that,
generally speaking, 
conspiracy information 
triggers larger cascades, 
gets more users involved, 
propagates `deeper',
persists longer, 
and is more viral and bursty than science information 
(Figure \ref{fig:spread_diff} A-F; $p\sim0$ for all Kolmogorov-Smirnov (K-S) tests\footnote{Kolmogorov-Smirnov test is abbreviated as K-S test hereafter.}; see also Table \ref{tab:ks_test_dynamics} in Appendix for details).

Specifically,
nearly 60\% of science posts 
% receive no any comments 
receive no comment
(i.e., cascade size is one), while only about 29\% of conspiracy posts do so.
More than 24\% of conspiracy cascades 
grow into more than the size of 10, 
but only less than 9\% of science cascades do so
(Figure \ref{fig:spread_diff}A).
However, we also find that only 0.035\% of conspiracy cascades (roughly 1 out of every 3,000 cascades)
have a size of more than 1000,
while 0.264\% of science cascades
(roughly 1 out of every 400 cascades)
do so
(Figure \ref{fig:spread_diff}A).
Similar patterns are also found in terms of number of unique users involved during the growth of cascades (Figure \ref{fig:spread_diff}B).
Moreover,
conspiracy information 
% literally 
is found to
propagate deeper (Figure \ref{fig:spread_diff}C) 
and be more viral (Figure \ref{fig:spread_diff}D) than science information.
For example, more than 15\% of conspiracy cascades 
go beyond a depth of 5, but less than 5\% of science cascades do so.
% 
% 
% Moreover,
The majority of conspiracy cascades also 
have a longer lifetime
% preserves longer 
than science cascades (Figure \ref{fig:spread_diff}E).
We also find that
conspiracy cascades
are slightly more bursty than science cascades (Figure \ref{fig:spread_diff}F),
suggesting
% that
% conspiracy cascades are formed in 
% shorter timeframes 
% than science cascades.

\begin{figure*}[!hb]
    \centering
    \includegraphics[scale=.5]{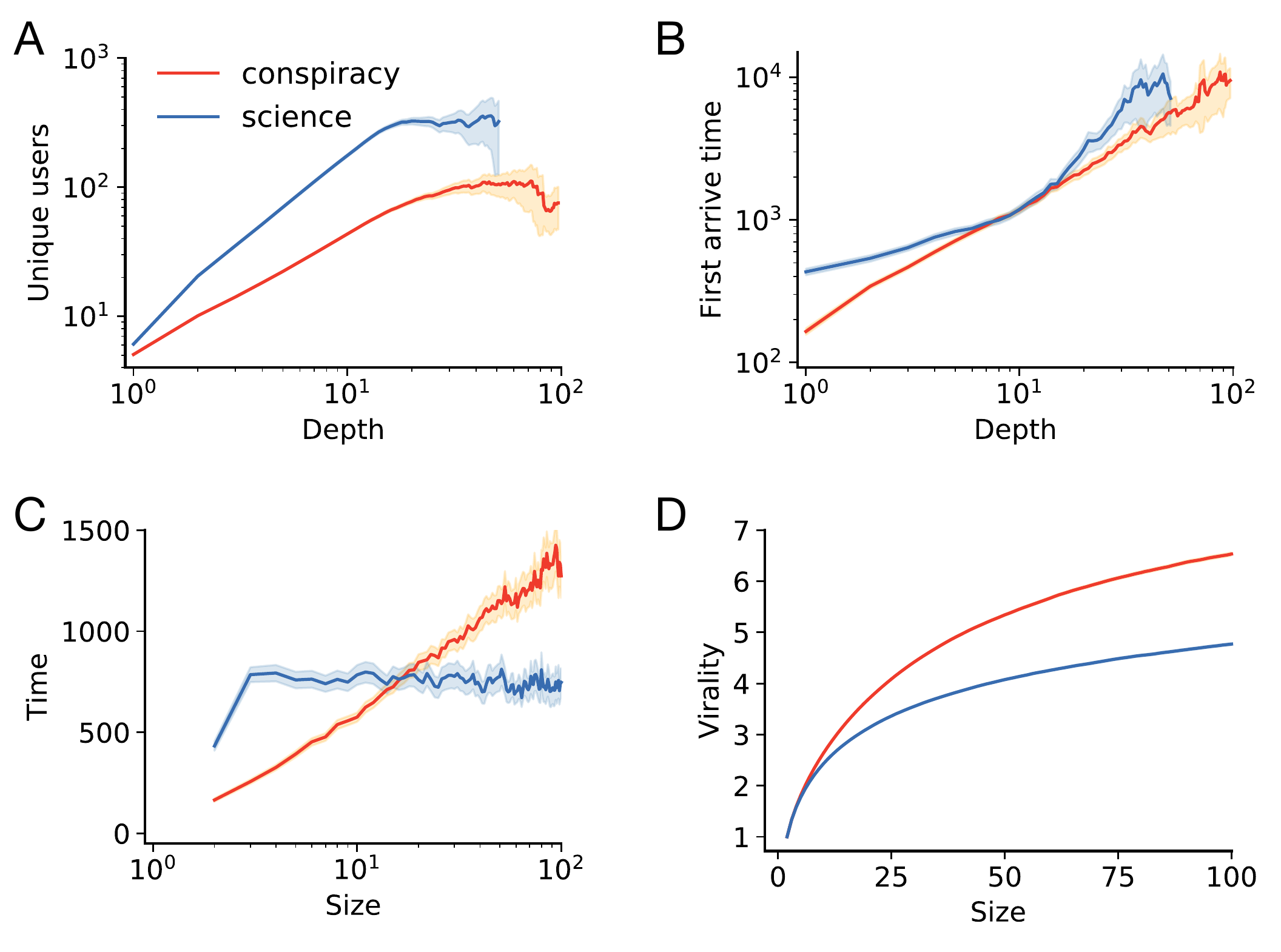}
    \caption{
    \textsf{
    % \small
    \footnotesize
    \textbf{Analysis of the growth of cascades.
    }
    (\textit{A}) Number of unique users reached at every depth.
    (\textit{B-C}) The time (in minutes) it takes for conspiracy and science cascades to reach any (B) depth and (C) size.
    (\textit{D}) The virality of cascades when conspiracy and science cascades reach any size.
    For ease of computation and visualization, cascade depth and size beyond 100 during the growth process are not shown.
    Shading areas indicate the 95\% Confidence Intervals (CIs).
    }}
    \label{fig:depth_size}
\end{figure*}

\noindent
% \ref{fig:spread_diff}F),
% suggesting 
that
conspiracy cascades are formed in 
shorter timeframes 
than science cascades.

With a close look at the growth process of cascades, we find that when cascade depth is controlled, science cascades tend to reach more users than conspiracy cascades (Figure \ref{fig:depth_size}A),
which further corroborates that science cascades tend to grow in a breadth-first manner compared with conspiracy cascades.
As expected,
it takes less time for conspiracy cascades to reach the same depth as science cascades (Figure \ref{fig:depth_size}B).
When cascade size is controlled, we find mixed patterns of the dynamic growth of cascades:
it takes conspiracy cascades less time to reach small sizes (e.g., size=5)
and much more time to reach relatively large sizes (e.g., size=50) (Figure \ref{fig:depth_size}C).
More importantly,
for the same cascade size, conspiracy cascades tend to be more viral than science cascades (Figure \ref{fig:depth_size}D).
% 
% In other words,
That said, conspiracy cascades are more likely to grow in a multigenerational branching manner compared with science cascades.

\begin{figure*}[!b]
    \centering
    \includegraphics[scale=.5]{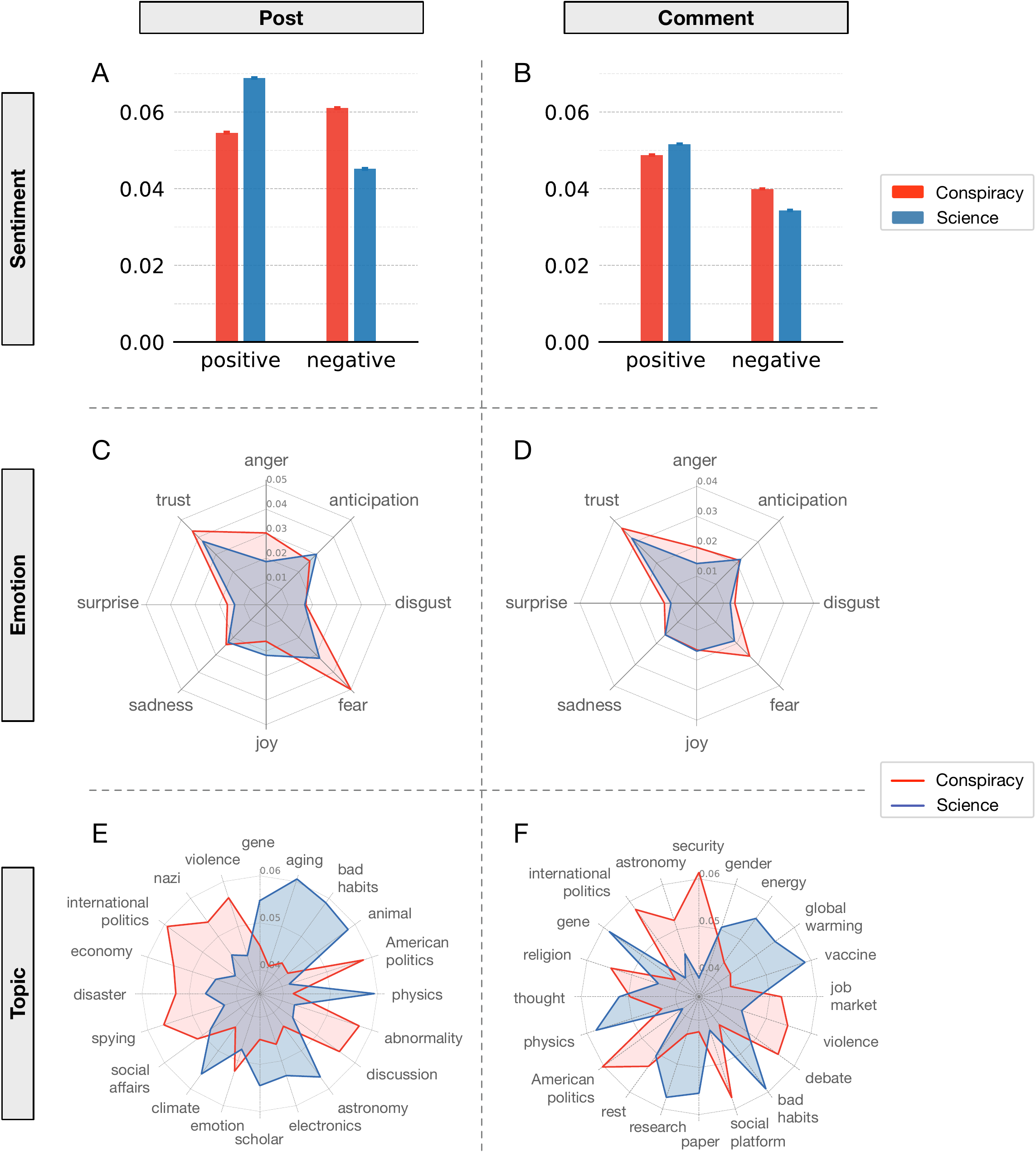}
    \caption{
    \textsf{
    % \small
    \footnotesize
    \textbf{Sentiment, emotion and topic distributions.
    % of posts and comments.
    }
    (\textit{A-B}) Sentiment distributions in posts and comments.
    Error bars indicate standard error of the mean.
    (\textit{C-D}) Emotion distributions in posts and comments.
    (\textit{E-F}) Topic distributions in posts and comments.
    }}
    \label{fig:emotion_topic}
\end{figure*}

\section{Content analysis}

Contents embedded in the cascades are thought to provide 
% additive power 
additional ingredients
to characterize the propagation of information.
For instance, 
negative moods can stimulate online trolling behaviors \cite{cheng2017anyone};
emotional words are able to catalyze the diffusion of messages \cite{brady2017emotion};
and political news is more likely to 
be retweeted
than general news \cite{vosoughi2018spread}.
Here we 
% aim to investigate the differences of
probe into
sentiments, emotions and topics conveyed in posts and comments of conspiracy and science cascades.

\subsection{Sentiment and emotion analysis}

% First,
To elicit the sentiments and emotions contained in posts and comments,
we adopt a leading word-emotion association lexicon managed by National Research Council Canada (NRC) \cite{mohammad2010emotions,mohammad2013crowdsourcing}.
This lexicon covers a manually annotated list of 14,182 words and their associations with
two sentiments (positive and negative) and
eight emotions (anger, anticipation, disgust, fear, joy, sadness, surprise and trust).
Based on the lexicon, the distributions of
each post and comment
over the two sentimental and eight emotional dimensions
are then calculated by the word frequency and further normalized by the content length.
% calculated and aggregated.
% 
For example, if a 
ten-word
post contains one positive word, three negative words, one word of anger, two words of fear and two words of trust,
the acquired sentimental and emotional dictionaries are then depicted as
$\{{\verb|positive|}: 1/10, {\verb|negative|}: 3/10\}$
and
$\{{\verb|anger|}: 1/10, {\verb|fear|}: 2/10, {\verb|trust|}:2/10\}$,
respectively.

\subsubsection{Sentiment}

Figure \ref{fig:emotion_topic}A illustrates the sentiment distribution of conspiracy and science posts, and Figure \ref{fig:emotion_topic}B shows the sentiment distribution of comments under these posts.
As we can see,
conspiracy posts contain far more negative words than science posts (K-S test = 0.165, $p \sim 0.0$),
where,
on average, 
a 100-word conspiracy post contains $\sim$6.1 negative words while a 100-word science post contains only $\sim$4.5 negative words (Figure \ref{fig:emotion_topic}A).
As expected, 
% Although 
conspiracy posts contain less positive words than science posts (K-S test = 0.077, $p \sim 0.0$) (Figure \ref{fig:emotion_topic}A).
Similarly, conspiracy posts inspire more negative (K-S test = 0.054, $p \sim 0.0$) and less positive (K-S test = 0.049, $p \sim 0.0$) contents than science posts in their comments (Figure \ref{fig:emotion_topic}B).

\subsubsection{Emotion}

The emotion distributions
of
posts and comments
in
conspiracy and science cascades
over eight dimensions 
are shown in Figure \ref{fig:emotion_topic} C and D.
At the macro level, conspiracy cascades convey more emotional contents than science cascades in both posts and comments,
despite the fact that the emotional gap
(indicated by the difference of the shading areas)
is larger in posts than in comments.
At the micro level, 
conspiracy cascades consistently convey more emotional contents of anger, fear, disgust and surprise
% and less emotional contents of joy 
than science cascades in both posts and comments
($p \sim 0.0$ for all K-S tests).
To our surprise,
% More interestingly,
conspiracy cascades also contain more contents of trust than science cascades ($p \sim 0.0$ for all K-S tests).
With a close look at the raw contents,
we find that 
many conspiracy posts are apt to use phrases like
``the truth about something" 
which appeal to the public and thereby would contribute to getting more traffics they need.
% 
% 
% Taken together,
In short,
the higher level of emotions expressed in conspiracy posts and comments may 
inspire more people to fuel 
the circulation of conspiracy information
than that of science information.

\subsection{Topic analysis}

To extract the topical concerns from contents, 
we adopt Latent Dirichlet Allocation (LDA) \cite{blei2003latent,phan2007gibbslda++},
a widely used method for topic modeling in the literature \cite{guan2016segmenting,park2018strength,singer2017why,vosoughi2018spread,way2016gender,zhang2017using},
in current study.
% Specifically,
% 
We implement two topic models with both 20 topics by LDA,
one for posts and another for comments.
Note that, for ease of presentation we set the number of topics as 20, but other reasonable choices, such as 50 or 100, would present similar results.
In practice,
we first remove stop words, punctuations, URLs, and retain only top 10,000 words 
according to their tf-idf (term frequency-inverse document frequency) weights
in posts or comments.
The remaining corpora are then fed into LDA topic models.
After that we obtain a probability distribution over 20 topics for each post or comment under the respective topic model.

Figure \ref{fig:emotion_topic} E and F illustrate the topical concerns
of
posts and comments 
in
conspiracy and science cascades
over 20 dimensions.
As shown in the figure, 
conspiracy and science cascades 
% focus on 
occupy
quite different topic spaces
($p \sim 0.0$ for all K-S tests in 
all paired comparisons between conspiracy and science cascades).
For example,
conspiracy posts 
are keen on
topics like politics, disaster, violence and economy,
while science posts mainly focus on socio-scientific related topics, such as gene, climate and physics (Figure \ref{fig:emotion_topic}E).
Similar patterns are also found in comments, as conspiracy comments 
% mainly focus on 
are keen on
topics like politics, security, violence and religion, while science comments 
still focus on socio-scientific related issues,
% mainly focus on social and health issues, 
including energy, global warming, vaccine and bad habits
(Figure \ref{fig:emotion_topic}F).
The clear topic differences
between 
conspiracy and science cascades
also imply that topical concerns 
are of potential utility to distinguish conspiracy cascades from science ones.

\section{Cascade classification}

To validate the utility of the described features above,
we consider a simple binary prediction task 
to determine whether a given post belongs to
conspiracy or science category.
We adopt Random Forest as the classifier due to its high performance in classification tasks.
In practice, 
We use the implementation from the {\verb|scikit-learn|}\footnote{\url{https://scikit-learn.org/stable/modules/generated/sklearn.ensemble.RandomForestClassifier.html}.} Python package with a forest of 200 trees.

\subsection{Feature sets}
% \vspace{2mm}
% \noindent
% \textbf{Feature sets:}
There are mainly three kinds of features that we would like to consider:
\begin{itemize}
    \item {\verb|dynamic|}: cascade dynamics, including cascade size, depth, virality, lifetime and number of unique users involved.
    \item {\verb|emotion|}: sentiments and emotions conveyed in the posts, including two sentiments (positive and negative) and eight emotions (anger, anticipation, disgust, fear, joy, sadness, surprise and trust).
    \item {\verb|topic|}: topics embedded in the posts over 20 dimensions.
\end{itemize}

\noindent
For the three sets of features, 
there are seven combinations of them in total.
For example, {\verb|topic|} indicates that only topic features are used for the classification task, while {\verb|all/topic|} indicates that all sets of features, except for topic features, are used for classification.

\begin{figure}[!htb]
    \centering
    \includegraphics[scale=.75]{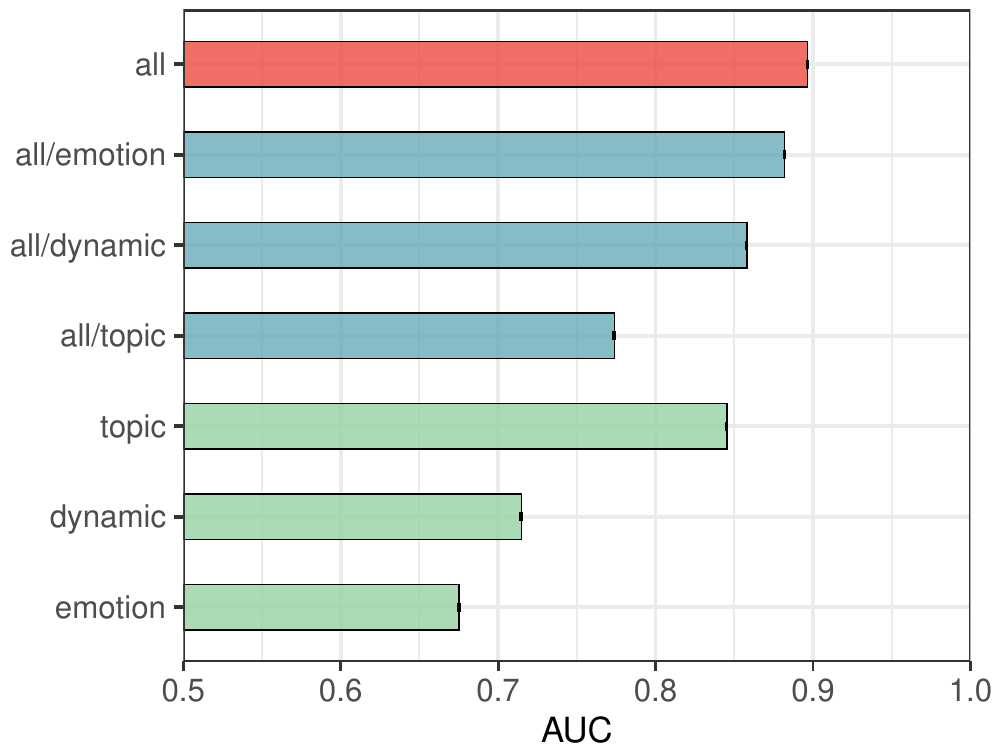}
    \caption{
    \textsf{
    % \small
    \footnotesize
    \textbf{Classification performance using the obtained features.}
    Error bars indicate the standard errors of the average AUCs obtained from ten random training-test dataset splits.
    }}
    \label{fig:clf_report}
\end{figure}

\subsection{Classification results}
For the classification task, 
there are 731,882 conspiracy and 734,327 science posts in total.
We randomly split the dataset into training (80\%) and test (20\%) sets.
The random forest classifier is first trained on the training set and then tested on the test set.
Figure \ref{fig:clf_report} presents the prediction performance in terms of the AUC (area under the curve) score 
% for all cascades 
on the test set
using any combination of feature sets as the predictors in ten random trials.
Note that the baseline--random guessing--would obtain an AUC score of 0.5 in our setting.
As such, our approach achieves very strong performance, with an AUC score of nearly 0.9
using all three sets of features.

As we can see from the figure, although each feature set alone significantly outperforms random guessing in the prediction task, 
it's the topic features that provide the most prominent predictive power (AUC=0.845), 
% topic features emerge as the the most prominent predictors,
followed by dynamic features (AUC=0.715) and emotion features (AUC=0.675).
In addition, 
the combination of topic features with dynamic features achieves an AUC score of 0.882,
and the combination of topic features with emotion features achieves an AUC score of 0.858.
However,
the combination of dynamic features with emotion features only achieves an AUC score of 0.774, 
which is even worse than 
% the AUC score achieved by 
the performance of
topic features alone.
This also implies that
topical concerns could be the primary factors that
% initiates the distinct spreading patterns of 
discriminate
conspiracy information from science information.
Taken together,
the classification results further favor the effectiveness of the proposed cascade features in depicting the differences between conspiracy and science narratives.

\section{Community participation}

\begin{table}[!b] \centering
   \caption{\textbf{Comment network description.}
    }
   \label{tab:comment_net_desp}
% Comment network basic description
% \begin{tabular}{lr@{\hskip .3in}lr|lr}
\begin{tabular}{lrrrrr}
% \toprule
\hline\noalign{\smallskip}
% \\[-3.2ex]
subreddit &   \#nodes & \#edges &   $<k>$ & $k_{in}^{max}$ & $k_{out}^{max}$ \\
% \\[-3.2ex]
% \midrule
\noalign{\smallskip}\hline\noalign{\smallskip}
conspiracy &   344,311 & 5,120,940 & 29.746 & 10,304 & 16,290 \\
science &   922,189 & 4,754,843 & 10.312 & 7,689 & 55,491 \\
% \bottomrule
\noalign{\smallskip}\hline
\end{tabular}
\end{table}

\begin{figure}[!t]
    \centering
    \includegraphics[scale=.5]{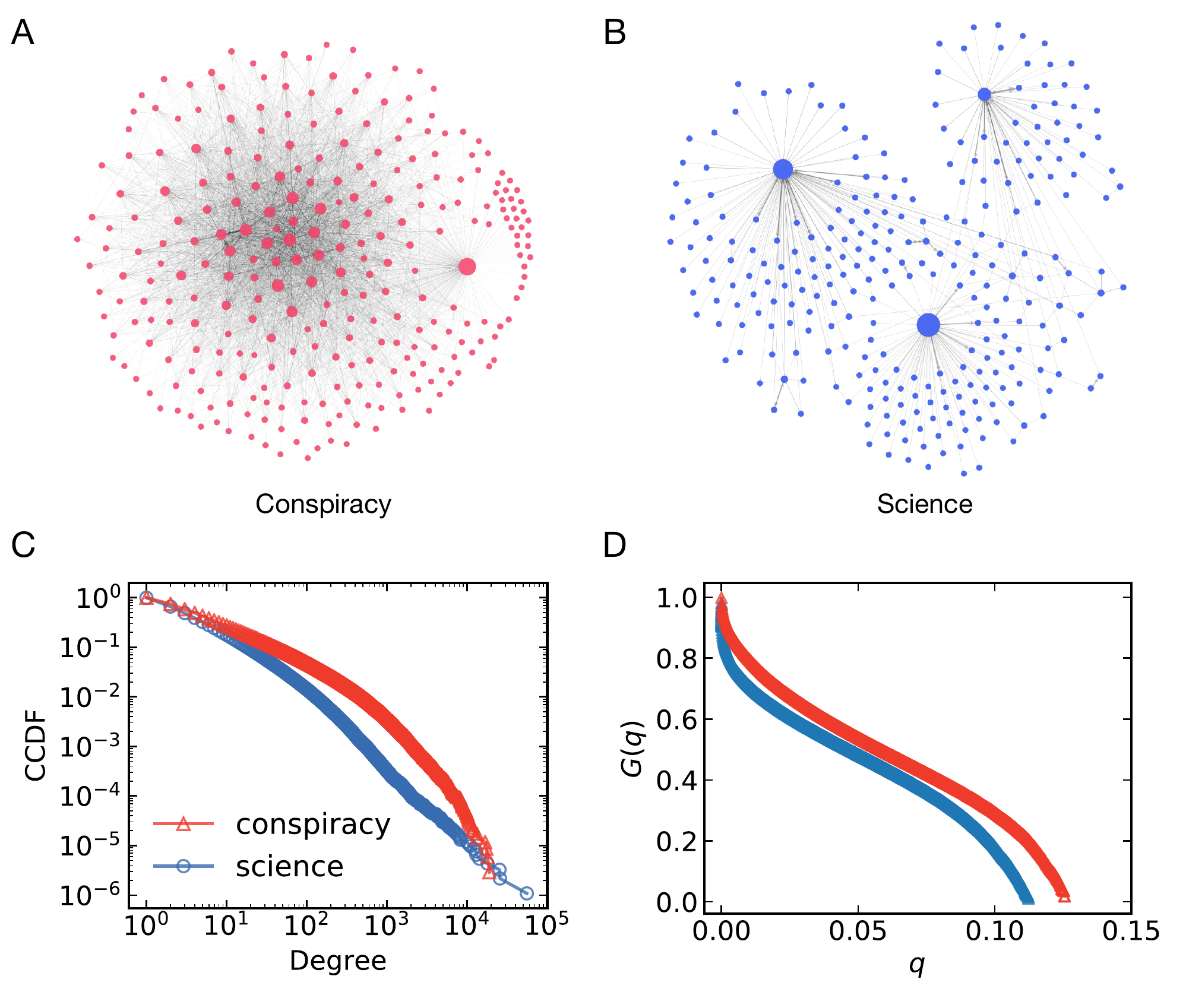}
    \caption{
    \textsf{
    % \small
    \footnotesize
    \textbf{Comment network.}
    (\textit{A}) Subnetwork extracted from conspiracy comment network.
    % with 276 nodes.
    (\textit{B}) Subnetwork extracted from science comment network.
    % with 274 nodes.
    (\textit{C}) Degree distributions of comment networks.
    (\textit{D}) Relative size $G(q)$ of the giant connected component as a function of the fraction of removed nodes $q$ in comment networks.
    }}
    \label{fig:comment_net}
\end{figure}

In this section, we examine the engagement of users 
during the propagation of conspiracy and science narratives.
We find that conspiracy cascades are less likely to be concentrated on focal users compared with science cascades.

To do so, we first construct two comment networks--one for conspiracy community and another for science community--where if user A leaves a comment to user B, there will be a directed link from B to A.
In practice,
users whom are hard to be identified (e.g., users denoted as `\textit{deleted}') are neglected during the construction of comment networks.
Figure \ref{fig:comment_net} A and B illustrate
two subnetworks that are randomly drawn from conspiracy and science comment networks, respectively.
Intuitively, conspiracy comment network is more densely connected than science comment network.
The degree (note that degree equals to the sum of indegree and outdegree for nodes in directed networks) distributions of nodes in conspiracy and science comment networks are shown in Figure \ref{fig:comment_net}C.
As shown in the figure,
% We can see that 
nodes in the conspiracy comment network tend to have higher degrees than nodes in the science comment network,
suggesting that users engaged in the circulation of conspiracy cascades 
% tend to 
are likely to
interact with more other users than 
those of users in science cascades.
% users engaged in science cascades.
% 
More detailed descriptions about conspiracy and science comment networks can be found in Table \ref{tab:comment_net_desp}.

To investigate to what extent the comment network is concentrated
% and the robustness of
and the possible strategies for mitigating the propagation process,
we implement a simple percolation process \cite{callaway2000network} on conspiracy and science comment networks respectively,
where node with the largest degree is removed at each step until the 
network is disconnected or no large component exists.
Figure \ref{fig:comment_net}D illustrates the 
relative size of the giant connected component (measured by size of the giant connected component divided by the original network size) as 
nodes are removed step by step during the percolation process.
After the removal of less than 15\% of nodes, both conspiracy and science comment networks are collapsed as the relative size of the giant connected component is less than 0.01 compared with the original network size.
However, science comment network is much more fragmented than conspiracy comment network for the same fraction of node removal,
which indicates that
user engagement in science cascades is more concentrated than that of conspiracy cascades.
In other words,
conspiracy cascades are less likely to be driven by a few focal users compared with science cascades.
This also hints that
to prevent the circulation of conspiracy narratives on the platform,
simply blocking some focal users or a small set of influencers may be
less effective than 
the same procedure implemented for science narratives.

\section{Political orientation}

Political affinity 
% of the information source
is thought to 
% have the potential to
play a role in
the circulation of misinformation
as people of different political orientations tend to consume specific kinds of misinformation
\cite{allcott2017social,bovet2019influence,grinberg2019fake,guess2018selective}.
Here we examine whether 
the political orientation of the 
% low-credibility contents 
information source
matters
% the findings still apply 
in the circulation of conspiracy narratives.

Most conspiracy posts contain url links in their contents, providing a feasible way to track where the information comes from.
Based on the affiliated url links (if any) and a classification of media outlets
curated by a
fact-checking organization\footnote{\url{https://mediabiasfactcheck.com}.},
we are able to assign posts to different political subgroups.
An example list of media outlets and their political orientations is given in Table \ref{tab:media_bias} in Appendix.
Specifically, 
% for a given post, if the
the information sources are sorted into several subgroups in three main categories:
($i$) social media sites, including Twitter, Facebook, Imgur, Google and Youtube;
($ii$) media outlets, including media outlets with left (e.g., MSNBC), left leaning (e.g., the New York Times), center (e.g., Reuters), right leaning (e.g., the Wall Street Journal) and right (e.g., Fox News) political orientations;
($iii$) homegrown news, such as news without a url or from other subreddits or communties in Reddit.

\begin{figure*}[!t]
    \centering
    \includegraphics[scale=.5]{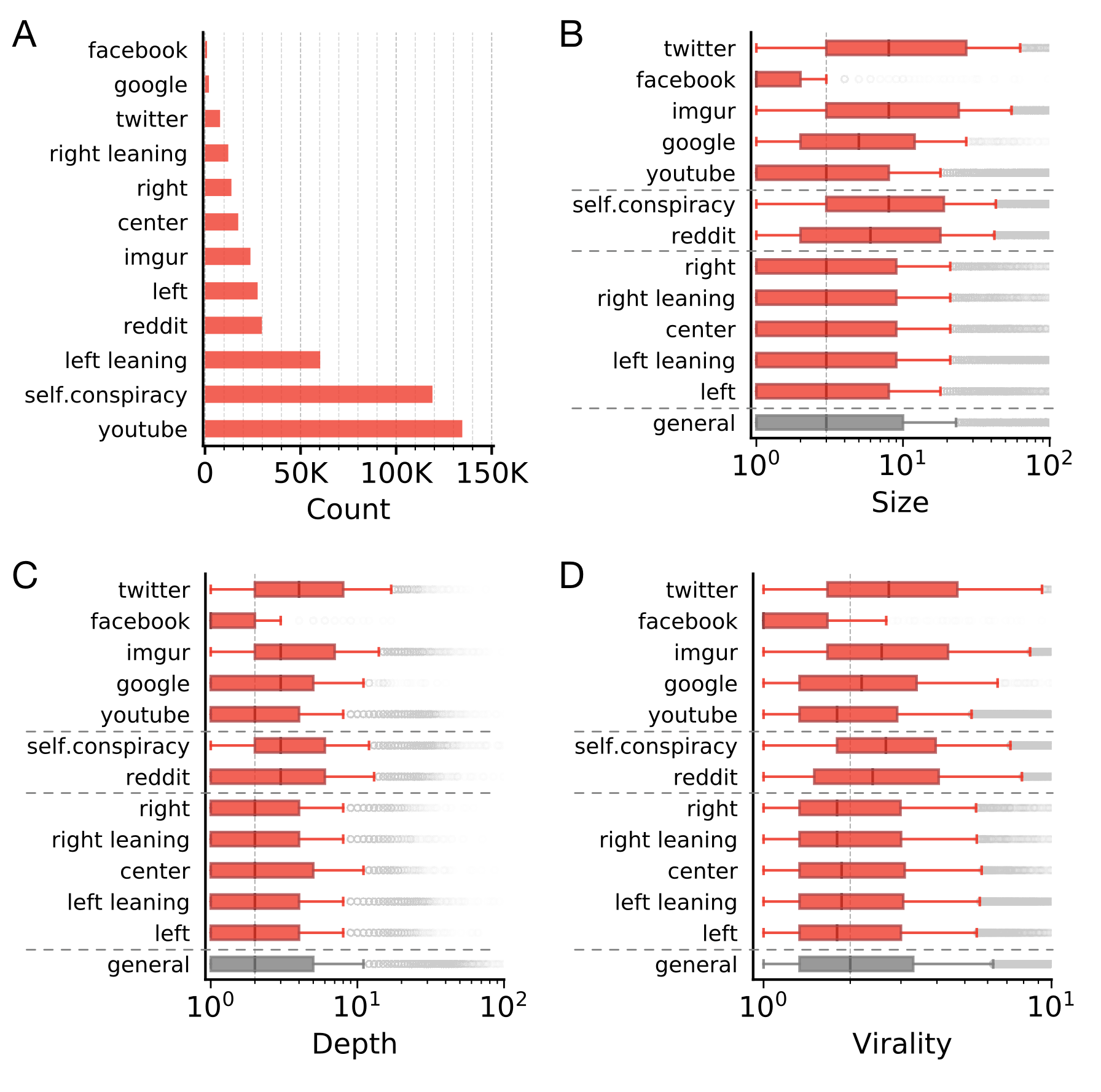}
    \caption{
    \textsf{
    % \small
    \footnotesize
    % \textbf{Post sources of different political orientations.}
    \textbf{Political orientation.}
    (\textit{A}) Cascade count by subgroups. 
    (\textit{B}) Cascade size by subgroups. 
    (\textit{C}) Cascade depth by subgroups. 
    (\textit{D}) Cascade virality by subgroups. 
    We show the box plot of cascade size, depth and virality in
    % (\textit{B}), (\textit{C}) and (\textit{D}),
    (\textit{B-D}),
    where the line inside each box indicates the median value and the grey circles indicate the outliers of the corresponding data.
    For ease of comparison, cascade size, depth and virality of all conspiracy cascades are shown in grey (denoted as \textit{general}),
    and the median values are shown in vertical dashed lines in
    (\textit{B-D}).
    }}
    \label{fig:political}
\end{figure*}

We restrict our analysis
to conspiracy posts that either
contain a url pointing to domains of the defined subgroups
or
have no
url in the content.
Figure \ref{fig:political}A illustrates the amounts of posts from different subgroups, where \textit{self.conspiracy} indicates posts without any url inside.
We can see that left and left leaning media outlets 
are more prevalent 
than right and right leaning media outlets in the conspiracy community.
We further compare the cascade size, depth and virality for different subgroups (Figure \ref{fig:political} B-D).
For ease of visualization,
only cascades 
whose depth and virality are 
greater than or equal to 1
% larger than or equal to 1
are considered in the analysis of cascade depth and virality (Figure \ref{fig:political} C-D),
but this doesn't alter the results 
as 
the conclusions still hold true
even after all cascades are included in the analysis.
For ease of comparison, 
we also show the corresponding quantities for the general case,
where cascade size, depth and virality for the complete conspiracy cascades
are shown in grey and denoted as \textit{general} in Figure \ref{fig:political} B-D.

As shown in Figure \ref{fig:political}B, the size of cascades
% from subgroups of
from different political subgroups, such as left, left leaning, right and right leaning,
is nearly the same with the general case, 
indicating that political orientation has very few effects in 
% determining 
driving
the ultimate cascade size.
Instead, we find that
posts from online social media sites like Twitter and Imgur
trigger larger cascades than the general,
with the exceptions of Facebook and Youtube.
Similar patterns are also found in terms of cascade depth (Figure \ref{fig:political}C).
We also find that cascades from
different political subgroups
are slightly less viral than the general case,
but cascades from online social media sites are generally more viral than other cascades (Figure \ref{fig:political}D).

Taken together, we find very 
% few 
little
evidence that political orientation of the information source
contributes to
the circulation of conspiracy narratives in current study.
Instead, we find that posts from social media sites with unclear political orientations generally trigger larger cascades and are more viral than the general.

\section{Discussion}

The over-proliferation of 
misinformation
online
could spark ``digital wildfires" in our hyperconnected world \cite{howell2013digital}
and
has triggered heated public debate in the aftermath of the 2016 US presidential election \cite{allcott2017social,bovet2019influence,grinberg2019fake}.
Our study presents a large-scale quantitative analysis
toward the understanding of the formation and growth of discussion cascades in online communities.
Specifically, we analyze the
cascade dynamics and contents of two distinct narratives:
conspiracy theories, whose claims are generally unsubstantiated,
and scientific information, whose origins are largely verifiable.
The results reveal systematic differences between conspiracy and science narratives in terms of cascade dynamics and contents.
Through a serial of classification tasks,
we show that topic concerns act as the primary factors that 
discriminate conspiracy narratives from science ones,
followed by dynamic and emotion features.
After the implementation of a simple percolation process on comment networks,
we find that 
the circulation of scientific information, rather than conspiracy theories,
relies more on focal users.
With a close look at the origins of conspiracy narratives,
we also find that, generally speaking,
posts from social media sites with unclear political orientations, instead of posts from left- or right-wing media outlets, trigger larger cascades and are more viral than the general.

Our study has practical implications for current social media and misinformation research.
First, 
as unsubstantiated news tends to focus on political and violent topics,
it's anticipated that future regulations on misinformation should pay close attention to such areas.
Second, 
as we have shown in the main text,
conspiracy narratives
are less likely to be driven by focal users compared with scientific information,
implying that, to prevent the dissemination of misinformation, simply removing or blocking some participants 
during the diffusion process
may be less effective than expected.
Third, as news from social media sites generally triggers larger cascades and is more viral than the general,
it's worth pointing out that 
% suggesting that 
social media sites could be important sources of misinformation,
thereby
highlighting
the urgent needs of proper regulations 
to suppress or mitigate the fabrication and dissemination of misinformation on social media sites.

The present work
has several limitations as well.
Our results are the outputs of one study conducted on two representative communities from Reddit,
but additional studies are urgently needed 
to validate and generalize our findings in other kinds of online communities or social domains
(such as \cite{del2016spreading,vosoughi2018spread}).
Our study is
based on descriptive and statistical analysis of observational data,
but confounds in the data may more or less skew the results.
For instance,
norms and incentives for how posts are submitted in these communities may be different, 
% thereby 
which could also
influence how information propagates in specific communities. 
As such,
future studies may follow rigorous causal inference approaches \cite{Aral858} (e.g., controlled experiments) 
to 
probe the social 
factors that catalyze the diffusion of misinformation and
elicit causal effects of misinformation on various social outcomes like
distress and elections.
In this comparative study, we have focused on the possibility of verification--instead of the quality--of conspiracy and science information.
Future studies may take the quality of information source into consideration.
In addition,
current study doesn't completely rule out the 
the role that
social bots play in the circulation of information.
But
to what extent is social diffusion vulnerable to social bots
is still an open and important question, 
and needs more comprehensive explorations and discussions
in the future.

% \section{Declarations}

% \textbf{Funding:} This work was supported by GRF 11505119 from Hong Kong RGC, CCR 9360120 and HKIDS 9360163 from City University of Hong Kong. 
% This work was also supported by the National Natural Science Foundation of China under Grant Nos 61773255, 61873167, 61473189, the Natural Science Foundation of Shanghai (No. 17ZR1445200), and the Science Fund for Creative Research Groups of the National Natural Science Foundation of China (No. 61521063).

% \vspace{2mm}
% \noindent
% \textbf{Competing interests:} The authors declare no competing interests.

% \vspace{2mm}
% \noindent
% \textbf{Availability of data and material:} The raw data we used in this study are acquired from and publicly available at \url{https://files.pushshift.io/reddit}.
% All other data are available from the corresponding authors on reasonable request.

% \vspace{2mm}
% \noindent
% \textbf{Code availability:} All scripts used for data processing and analysis are available from the corresponding authors on reasonable request.

% \vspace{2mm}
% \noindent
% \textbf{Authors' contributions:} Y.Z., L.W., J.J.H.Z., and X.W. conceived the present study. Y.Z. processed and analyzed the data. Y.Z. wrote the paper with input from L.W., J.J.H.Z., and X.W.

\vspace{5mm}

\begin{acknowledgements}
This work was supported by the National Natural Science Foundation of China (Grant Nos. 61773255 and 61873167), Hong Kong RGC (GRF 11505119) and City University of Hong Kong (CCR 9360120 and HKIDS 9360163). The authors would like to thank Tai-Quan ``Winson" Peng for critical reading of the early draft.
% This work was supported by GRF 11505119 from Hong Kong RGC, CCR 9360120 and HKIDS 9360163 from City University of Hong Kong. 
% This work was also supported by the National Natural Science Foundation of China under Grant Nos 61773255, 61873167, 61473189, the Natural Science Foundation of Shanghai (No. 17ZR1445200), and the Science Fund for Creative Research Groups of the National Natural Science Foundation of China (No. 61521063).
% The authors would like to thank 
% Tai-Quan ``Winson" Peng
% for critical reading of the early draft.

\end{acknowledgements}


% \bibliographystyle{ACM-Reference-Format}
\bibliographystyle{spmpsci}      % basic style, author-year citations
\bibliography{sample}

\newpage

% \appendix

\noindent \textbf{Appendix}

\begin{table}[!htbp] \centering
   \caption{\textbf{Data descriptions.}
   When calculating the mean values, quantities with highly skewed distribitions are log-transformed by $log_{10}(x+1)$.
    }
   \label{tab:ks_test_dynamics}

\setlength{\tabcolsep}{3.8pt}

Cascade size (KS-test: $D=0.34323006611689794$, $p\sim0.0$)
\begin{tabular}{lrrrrrrrr}
\toprule
% \hline\noalign{\smallskip}
% \hline
  subreddit &     count &    mean (log) &      std (log) &  min &  50\% &  max \\
\midrule
% \noalign{\smallskip}\hline\noalign{\smallskip}
 conspiracy &  731882 &  0.776 &   0.487 &  1 &   3 &   4617 \\
    science &  734327 &  0.515 &  0.425 &  1 &   1 &   12412 \\
\bottomrule
% \hline\noalign{\smallskip}
% \noalign{\smallskip}\hline
\end{tabular}

\vspace{8mm}

Unique users (KS-test: $D=0.3331162726782291$, $p\sim0.0$)\\
\setlength{\tabcolsep}{3.8pt}
\begin{tabular}{lrrrrrrrr}
\toprule
% \hline\noalign{\smallskip}
  subreddit &     count &    mean (log) &      std (log) &  min &  50\% &      max \\
\midrule
% \noalign{\smallskip}\hline\noalign{\smallskip}
 conspiracy &  731882 &  0.670 &  0.383 &  1 &  3 &  1698 \\
    science &  734327 &  0.469 &  0.348 &  1 &  1 &  5049 \\
\bottomrule
% \noalign{\smallskip}\hline
\end{tabular}

\vspace{8mm}

Depth (KS-test: $D=0.3074821114225904$, $p\sim0.0$)
\setlength{\tabcolsep}{3.7pt}
\begin{tabular}{lrrrrrrrr}
\toprule
% \hline\noalign{\smallskip}
  subreddit &     count &    mean (log) &      std (log) &  min &  50\% &      max \\
\midrule
% \noalign{\smallskip}\hline\noalign{\smallskip}
 conspiracy &  731882 &  0.418 &  0.357 &  0 &   1 &   429 \\
    science &  734327 &  0.190 &  0.279 &  0 &   0 &   1293 \\
\bottomrule
% \noalign{\smallskip}\hline
\end{tabular}

\vspace{8mm}

Virality (KS-test: $D=0.34323006611689794$, $p\sim0.0$)
\setlength{\tabcolsep}{2.7pt}
\begin{tabular}{lrrrrrrrr}
\toprule
% \hline\noalign{\smallskip}
  subreddit &     count &    mean (log) &      std (log) &  min &  50\% &      max \\
\midrule
% \noalign{\smallskip}\hline\noalign{\smallskip}
 conspiracy &  731882 &  0.369 &  0.285 &  0 &  1.333 &   163.156 \\
    science &  734327 &  0.176 &  0.239 &  0 &  0 &  462.819 \\
\bottomrule
% \noalign{\smallskip}\hline
\end{tabular}

% \noalign
% \begin{flushleft}
% \vspace{2mm}
% (Continued in next page)
% \end{flushleft}

\vspace{8mm}

\setlength{\tabcolsep}{12pt}
Lifetime (KS-test: $D=0.3767388086135375$, $p\sim0.0$)\\
\setlength{\tabcolsep}{2pt}
\begin{tabular}{lrrrrrrrr}
\toprule
% \hline\noalign{\smallskip}
  subreddit &     count &    mean (log) &      std (log) &  min &  50\% &      max \\
\midrule
% \noalign{\smallskip}\hline\noalign{\smallskip}
 conspiracy &  731882 &  1.872 &  1.410 &  0 &   189 &   259185 \\
    science &  734327 &  0.877 &  1.324 &  0 &  0 &  259195 \\
\bottomrule
% \noalign{\smallskip}\hline
\end{tabular}

\vspace{8mm}
\setlength{\tabcolsep}{5pt}
Burstiness (KS-test: $D=0.05736421439137$, $p\sim2.67e-280$)
\begin{tabular}{lrrrrrrrr}
\toprule
% \hline\noalign{\smallskip}
  subreddit &     count &   mean &    std &  min &    50\% &    max \\
\midrule
% \noalign{\smallskip}\hline\noalign{\smallskip}
 conspiracy &  364115 &  0.133 &  0.330 & -1.0 & 0.141 &   0.965 \\
    science &  133797 &  0.128 &  0.383 & -1.0 & 0.102 &   0.970 \\
\bottomrule
% \noalign{\smallskip}\hline
\end{tabular}

\end{table}

% \newpage

% \vspace{10mm}

\begin{table}[!htb] \centering
   \caption{\textbf{Media outlets and their political orientations.}
   We show the top 10 domain names under each category and the corresponding number ($N$) of posts found in the data.
  For space constrains, only categories of \textit{left}, \textit{left leaning}, \textit{right leaning} and \textit{right} are shown.
    }
   \label{tab:media_bias}

\setlength{\tabcolsep}{4.5pt}
% \begin{tabular}{lr@{\hskip .3in}lr|lr}
\begin{tabular}{lrclr}
\toprule
% \hline\noalign{\smallskip}
\\[-3.2ex]
\multicolumn{2}{c}{left} & \multicolumn{3}{c}{left leaning} \\
% & \multicolumn{3}{c}{center} \\ 
% \cline{1-2}
\\[-3.2ex]
\cmidrule(lr){1-2} \cmidrule(lr){4-5}
\\[-3.2ex]
             domain &   $N$ &&               domain &   $N$  \\
\\[-3.2ex]
\midrule
% \noalign{\smallskip}\hline\noalign{\smallskip}
 huffingtonpost.com &  2,577 &&      theguardian.com &  5,178  \\
            cnn.com &  2,175 &&          nytimes.com &  4,117  \\
       alternet.org &  1,684 &&   washingtonpost.com &  3,977  \\
       rawstory.com &  1,391 &&    independent.co.uk &  2,519  \\
          salon.com &  1,329 &&       news.yahoo.com &  1,849  \\
           wsws.org &  1,303 &&  businessinsider.com &  1,839  \\
   commondreams.org &  1,225 &&        bloomberg.com &  1,676  \\
   counterpunch.org &  1,122 &&          thehill.com &  1,615  \\
  thedailybeast.com &  1,117 &&            bbc.co.uk &  1,597  \\
   democracynow.org &   988 &&              bbc.com &  1,485  \\
\bottomrule
% \noalign{\smallskip}\hline
\end{tabular}

\vspace{8mm}

\setlength{\tabcolsep}{2pt}
\begin{tabular}{lrclr}
\toprule
% \hline\noalign{\smallskip}
\\[-3.2ex]
\multicolumn{2}{c}{right leaning} & \multicolumn{3}{c}{right} \\ 
\\[-3.2ex]
\cmidrule(lr){1-2} \cmidrule(lr){4-5}
\\[-3.2ex]
               domain &   $N$ &&                  domain &   $N$ \\
\midrule
% \noalign{\smallskip}\hline\noalign{\smallskip}
      telegraph.co.uk &  2,031 &&           breitbart.com &  2,341 \\
        wikileaks.org &  1,393 &&             foxnews.com &  2,038 \\
           nypost.com &  1,008 &&         dailycaller.com &  1,379 \\
  washingtontimes.com &   988 &&                 wnd.com &   821 \\
   russia-insider.com &   885 &&  washingtonexaminer.com &   702 \\
           forbes.com &   770 &&           express.co.uk &   585 \\
           reason.com &   492 &&            theblaze.com &   468 \\
     news.antiwar.com &   452 &&      thenewamerican.com &   467 \\
              wsj.com &   425 &&          freebeacon.com &   459 \\
 original.antiwar.com &   386 &&            thesun.co.uk &   309 \\
\bottomrule
% \noalign{\smallskip}\hline
\end{tabular}

\end{table}

\end{document}